\documentstyle[12pt]{article}

\title{
{\vspace{-3cm} \normalsize
\hfill \parbox{30mm}{DESY 94-088}}\\[25mm]
Numerical simulations and  \\
the strength of the electroweak phase transition   \\[8mm]}
\author{
{Z. Fodor}\thanks{On leave from Institute for Theoretical Physics,
E\"otv\"os University, Budapest, Hungary.},
J. Hein, K. Jansen, A. Jaster, I. Montvay  \\[4mm]
Deutsches Elektronen-Synchrotron DESY, \\
Notkestr.\,85, D-22603 Hamburg, Germany \\[8mm]
\and
F. Csikor  \\[4mm]
Institute for Theoretical Physics, E\"otv\"os University, \\
Budapest, Hungary}

\date{May 24, 1994}

\newcommand{\be}{\begin{equation}}
\newcommand{\ee}{\end{equation}}
\newcommand{\half}{\frac{1}{2}}

\begin{document}
\maketitle

\begin{abstract} \normalsize
Numerical simulations are performed to study the finite temperature
phase transition in the SU(2) Higgs model on the lattice.
The strength of the first order phase transition is investigated
by determining the latent heat and the interface tension on
$L_t=2$ lattices.
The values of the Higgs boson mass presently chosen are below 50 GeV.
Our results are in qualitative agreement with two-loop resummed
perturbation theory.
\end{abstract}
\vspace{1cm}


\section{Introduction}
 The standard calculational method for the study of the symmetry
 restoring electroweak phase transition \cite{KIRLIN} is resummed
 perturbation theory \cite{CARRIN,BUFOHEWA,ARNESP}.
 The highest known order in the gauge- and scalar quartic coupling is
 $g^4,\lambda^2$, combined with a high temperature expansion
 \cite{FODHEB}.
 In the Higgs phase with broken symmetry perturbation theory is
 expected to work well for not very high Higgs boson masses, since the
 couplings are small.
 The perturbative calculations can be extended to the vicinity of the
 symmetric phase for $\lambda \ll g^2$.
 In the high temperature symmetric phase, however, the situation is
 similar to high temperature QCD.
 Therefore, irreparable infrared singularities occur which prevent
 a quantitative control of graph resummation \cite{LINDE}.
 Since the calculation of physical quantities characterizing the
 phase transition requires the knowledge of both phases, there is
 a priory no reason why perturbation theory could provide a quantitative
 treatment of the electroweak phase transition for physically relevant
 couplings.

 An obvious non-perturbative method for the study of the symmetry
 restoring phase transition is numerical simulation on the lattice.
 Omitting, for simplicity, fermions and the U(1) gauge field one is
 left with the SU(2) Higgs model describing the interaction of
 a four-component Higgs scalar field with the SU(2) gauge field.
 After pioneering works \cite{DAMHEL,EVJEKA}, recent numerical
 simulations contributed a lot to the understanding of the finite
 temperature behaviour of the SU(2) Higgs model \cite{BIKS}.
 Another line of research has also been initiated recently
 \cite{KARUSH,JAKAPA}.
 In this approach one studies the three-dimensional effective
 theory obtained by dimensional reduction in the high-temperature
 limit.

 An important issue is the order and strength of the phase transition.
 For instance, the possibility of electroweak baryogenesis
 \cite{KURUSH} requires a strong enough first order electroweak
 phase transition \cite{SHAPOSH} (for a recent review see
 \cite{COKANE}).
 Up to now the four dimensional numerical simulations \cite{BIKS} have
 been concentrated on relatively heavy Higgs boson masses,
 where the properties of the phase transition are technically
 difficult to disentangle.
 In the numerical approach based on dimensional reduction indications
 have been found that the phase transition is stronger than given by
 perturbation theory \cite{FAKARUSH}.
 In the present letter we investigate this question in the four
 dimensional SU(2) Higgs model for lighter Higgs masses, below
 50 GeV.
 Since this region of parameters of the minimal standard model is
 already excluded by experiments, our present scope is merely
 theoretical because we would like to check the validity of
 some other theoretical approximation schemes, e.g.\ resummed
 perturbation theory.
 We plan to extend this investigation to heavier Higgs boson masses
 in future papers.

 The present letter is our first short communication.
 Technical details will not be included here, but will be postponed
 to a subsequent publication with more numerical data \cite{FUTURE}.


\section{Physical parameters}
 The lattice action of the SU(2) Higgs model is conventionally
 written as
$$
S[U,\varphi] = \beta \sum_{pl}
\left( 1 - \frac{1}{2} {\rm Tr\,} U_{pl} \right)
$$
\be \label{eq01}
+ \sum_x \left\{ \half{\rm Tr\,}(\varphi_x^+\varphi_x) +
\lambda \left[ \half{\rm Tr\,}(\varphi_x^+\varphi_x) - 1 \right]^2 -
\kappa\sum_{\mu=1}^4
{\rm Tr\,}(\varphi^+_{x+\hat{\mu}}U_{x\mu}\varphi_x)
\right\} \ .
\ee
 Here $U_{x\mu}$ denotes the SU(2) gauge link variable and
 $\varphi_x$ is a complex $2 \otimes 2$ matrix in isospin space
 describing the Higgs scalar field and satisfying
\be \label{eq02}
\varphi_x^+ = \tau_2\varphi_x^T\tau_2 \ .
\ee
 The bare parameters in the action are $\beta \equiv 4/g^2$ for
 the gauge coupling, $\lambda$ for the scalar quartic coupling and
 $\kappa$ for the scalar hopping parameter related to the bare
 mass square $\mu_0^2$ by $\mu_0^2 = (1-2\lambda)/\kappa - 8$.

 In order to fix the physical parameters in a numerical simulation
 one has to define and compute some suitable renormalized quantities
 at zero temperature.
 The renormalized gauge coupling can be determined from the static
 potential of an external SU(2) charge pair, measured by Wilson loops.
 The physical Higgs mass $M_H$ can be extracted from correlation
 functions of quantities as the site variable
\be \label{eq03}
R_x \equiv \half{\rm Tr\,}(\varphi_x^+\varphi_x)
\equiv \rho_x^2 \ ,
\ee
 or, using $\varphi_x \equiv \rho_x \alpha_x$, the link variables
\be \label{eq04}
L_\varphi \equiv
\half {\rm Tr\,}(\varphi^+_{x+\hat{\mu}}U_{x\mu}\varphi_x) \ ,
\hspace{3em}
L_\alpha \equiv
\half {\rm Tr\,}(\alpha^+_{x+\hat{\mu}}U_{x\mu}\alpha_x) \ .
\ee
 The W-boson mass $M_W$ can be obtained similarly from the composite
 link fields ($r,k=1,2,3$)
\be \label{eq05}
W_{rk} \equiv
\half {\rm Tr\,}(\tau_r \alpha^+_{x+\hat{k}}U_{xk}\alpha_x) \ .
\ee

 Since we are interested in the study of the symmetry restoring
 phase transition on lattices with temporal extension $L_t=2$,
 we have to determine the renormalized parameters at critical
 points for the $L_t=2$ lattices.
 In order to have a renormalized gauge coupling near the physical
 value $g_R^2 \simeq 0.5$ we choose $\beta=8$ \cite{LAMOWE}.
 As stated before, we would like to have lighter Higgs boson masses
 than studied in \cite{BIKS}.
 Therefore we have chosen the two values $\lambda = 0.0001$
 (referred to as {\em low}) and $\lambda = 0.0005$ (referred to as
 {\em high}).
 In these cases the critical hopping parameters turned out to be
 near $\kappa_c^{low} = 0.1283$ and $\kappa_c^{high} = 0.1289$,
 respectively.
 For the numerical simulation of the zero temperature ($T=0$)
 system we took $12^3 \cdot 24$ lattices, which turned out to be
 large enough for the present $L_t=2$ simulations.
 (Runs on $16^3 \cdot 32$ are planned in near future \cite{FUTURE}.)

 The numerical Monte Carlo simulations have been performed on the
 Quadrics Q16 machines at DESY.
 In order to decrease autocorrelations we applied a mixture of
 heatbath and overrelaxation algorithms with an optimized ratio,
 which depends on lattice sizes and bare parameters.
 An important improvement could be achieved by an overrelaxation
 algorithm in the Higgs field length ($\rho$) variable which has been
 developed and tested by two of us \cite{FODJAN}.
 The heatbath algorithm for the gauge field can be taken over from
 pure gauge theory \cite{KENPEN}.
 For the Higgs field one can start a heatbath algorithm at our small
 $\lambda$ values by generating an optimized Gaussian part of the
 single spin distribution and then correct by an accept-reject step
 for the remaining quartic piece \cite{BUNK}.
 Away from the critical $\kappa$ such combination of updating sweeps,
 which we call "complete sweep", almost entirely removes the
 autocorrelation.
 On our $12^3 \cdot 24$ lattices the integrated autocorrelations
 for different characteristic quantities are between 2 and 20 complete
 sweeps.
 The smallest integrated autocorrelations are shown by Wilson loops,
 the largest ones by variables proportional to the length of the
 Higgs field.
 In the metastability region with tunneling between the two phases
 we still have the "supercritical slowing down", however, in the
 simulations presented here this does not play a r\^ole.
 In the two-coupling simulations of sections 3 and 4 the observed
 integrated autocorrelations were maximally of the order of 100 to 300
 complete sweeps.

 Performing 160000 complete sweeps on $12^3 \cdot 24$ lattices and
 using an exponential + constant fit for the timeslice correlations of
 the quantities in (\ref{eq03}-\ref{eq04}) one obtains the following
 masses in lattice units:
$$
aM_H^{low} = 0.236(7) \ ,
\hspace{3em}
aM_H^{high} = 0.262(9) \ ,
$$
\be \label{eq06}
aM_W^{low} = 1.059(24) \ ,
\hspace{3em}
aM_W^{high} = 0.427(8) \ .
\ee
 Here, as everywhere in this paper, statistical errors in last digits
 are given in parentheses.
 The estimates of the statistical errors always come from dividing our
 simulation data into independent subsets and performing the fits
 in these subsets in order to obtain the variance of the fit parameters.
 This procedure is particularly simple to implement on a Quadrics
 parallel computer, where in most cases several independent lattices
 are simulated.
 In our case we started L\"uscher's random number generator
 \cite{LUSCHRAN} differently on each node, leading to statistically
 independent samples on the independent lattices.

 In general we do not yet attempt to estimate our systematical errors,
 because they are presumably dominated by the present limitation to
 $L_t=2$ lattices for the high temperature simulations.
 In the continuum limit one has to consider $L_t \gg 1$, which is
 possible, but can be demanding \cite{FUTURE}.

 For the mass ratio
 $R_{HW} \equiv M_H/M_W \equiv \sqrt{8\lambda_R}/g_R$ (\ref{eq06})
 gives
\be \label{eq07}
R_{HW}^{low} = 0.222(12) \ ,
\hspace{3em}
R_{HW}^{high} = 0.614(32) \ .
\ee
 With $M_W \simeq 80$ GeV this means that the {\em low} point is at
 $M_H \simeq 18$ GeV and the {\em high} point at $M_H \simeq 49$ GeV.
 On the $L_t=2$ lattices we have $aT_c=0.5$, therefore (\ref{eq06})
 also implies
\be \label{eq08}
T_c^{low} = 0.472(11) \cdot M_W \ ,
\hspace{3em}
T_c^{high} = 1.171(22) \cdot M_W \ .
\ee

 The static potential has been extracted from $r \otimes t$ Wilson
 loops with $1 \le r \le 6$ and $1 \le t \le 12$ on $12^3 \cdot 24$.
 We determined every such Wilson loop after transforming the gauge
 configuration to temporal gauge.
 The $t$-dependence has been fitted by three exponentials in order
 to determine the large-$t$ asymptotics.
 The potential can be well fitted by a Yukawa-potential with
 lattice corrections, as discussed in \cite{LAMOWE}.
 Taking a form
\be \label{eq09}
V(r) = -\frac{A}{r} e^{-Mr}+C+D \cdot G(M,r) \ ,
\ee
 with $G(M,r)$ being the difference of lattice Yukawa-potential and
 continuum Yukawa potential, we obtained with $\chi^2$ of order one:
$$
A^{low} = 0.0336(3)   \ , \hspace{1em}
M^{low} = 1.112(15)   \ , \hspace{1em}
C^{low} = 0.066453(6) \ , \hspace{1em}
D^{low} = 0.0343(23)  \ ,
$$
\be \label{eq10}
A^{high} = 0.03434(7)   \ , \hspace{1em}
M^{high} = 0.4272(21)   \ , \hspace{1em}
C^{high} = 0.090768(18) \ , \hspace{1em}
D^{high} = 0.0352(8)  \ .
\ee

 The excellent agreement of the masses $M$ in (\ref{eq10}) with $M_W$
 in (\ref{eq06}) and the nearly equality of the fit parameters $A$ and
 $D$ show that the static potential is well understood, and can be
 used for the definition of the renormalized gauge coupling.
 One could take for a definition $g_R^2 = 16\pi A/3$, but we prefer
 here not to use a global fit.
 Instead, we adopt for the definition of the renormalized gauge
 coupling a procedure similar to the one recently proposed in pure
 gauge theory \cite{SOMMER}.
 Namely, starting from eq. (35) in ref. \cite{LAMOWE} we define
 an $r$-dependent renormalized gauge coupling, which we then interpolate
 to a distance $r = M_W^{-1}$.
 The result is, this time including also an estimate of the systematic
 errors as the last entry in parentheses:
\be \label{eq11}
g_R^{2,low} = 0.5476(5+75+10) \ ,
\hspace{3em}
g_R^{2,high} = 0.5781(4+88+3) \ .
\ee
 The first two entries are the statistical errors from the Wilson loops
 and $M_W$, respectively.
 As one can see, the errors are dominated by the statistical errors of
 $M_W$ used as input in this analysis.
 This can be easily improved by better statistics, if later on
 required.
 Otherwise the errors are quite small, and the renormalization
 compared to the bare value $g^2=0.5$ is moderate.


\section{Latent heat}
 The pressure ($P$) is continuous, therefore the latent heat
 (i.~e. the discontinuity of the energy density $\Delta\epsilon$) can
 be calculated from the discontinuity of the quantity
 $\delta \equiv \epsilon/3-P$.
 Since $\delta$ is given by \cite{MONMUN}
\be \label{eq12}
\delta = \frac{1}{3} (TL_t)^4
\left\langle \frac{\partial\kappa}{\partial\tau} \cdot 8L_\varphi
- \frac{\partial\lambda}{\partial\tau} \cdot Q_x
- \frac{\partial\beta}{\partial\tau} \cdot 6P_U \right\rangle \ ,
\ee
 where $\tau \equiv \log(aM_W)^{-1}$, and besides $L_\varphi$ in
 (\ref{eq04}) we used
\be \label{eq13}
Q_x \equiv (\rho_x^2 - 1)^2 \ ,
\hspace{3em}
P_U \equiv 1 - \half {\rm Tr\,} U_{pl} \ .
\ee
 The vacuum contribution is not subtracted in (\ref{eq12}), but in
 the latent heat it cancels and we obtain
\be \label{eq14}
\frac{\Delta\epsilon}{T_c^4} = L_t^4
\left\langle \frac{\partial\kappa}{\partial\tau} \cdot 8\Delta L_\varphi
- \frac{\partial\lambda}{\partial\tau} \cdot \Delta Q_x
- \frac{\partial\beta}{\partial\tau} \cdot 6\Delta P_U \right\rangle \ .
\ee

 The partial derivatives are taken here along the {\em lines of
 constant physics} (LCP's), which can be defined by keeping constant
 the value of the mass ratio $R_{HW}$ and the renormalized gauge
 coupling $g_R$.
 For weak bare couplings one can estimate the change
 of $g^2 = 4/\beta$ and $\lambda_0 \equiv \lambda/(4\kappa^2)$
 by integrating the one-loop renormalization group equations
$$
\frac{dg^2(\tau)}{d\tau} = \frac{1}{16\pi^2} \left[
- \frac{43}{3}g^4 + O(\lambda_0^3,\lambda_0^2g^2,\lambda_0g^4,g^6)
\right] \ ,
$$
\be \label{eq15}
\frac{d\lambda_0(\tau)}{d\tau} = \frac{1}{16\pi^2} \left[
96\lambda_0^2 + \frac{9}{32}g^4 - 9\lambda_0 g^2 +
O(\lambda_0^3,\lambda_0^2g^2,\lambda_0g^4,g^6) \right] \ .
\ee
 Starting from our {\em low} or {\em high} critical point of the
 $L_t=2$ lattice one can obtain the corresponding $L_t=3,4,5,\ldots$
 points by integrating (\ref{eq15}) from $\tau=0$ to
 $\tau = \log(3/2),\log(4/2),\log(5/2),\ldots$.
 For instance, the $L_t=3$ critical point corresponding to our {\em low}
 point ($\beta=8,\lambda=0.0001,\kappa=0.12830$) comes out to be at
 ($\beta=8.147,\lambda=0.000111$) and the one corresponding to the
 {\em high} point ($\beta=8,\lambda=0.0005,\kappa=0.12887$) at
 ($\beta=8.147,\lambda=0.000507$).

 Of course, we still need another equation for the change of the
 critical $\kappa = \kappa_c$.
 The critical hopping parameter can be determined by different methods.
 We found it convenient to apply a two-coupling method: on a long
 lattice in one spatial direction (say, $z$-direction) the lattice
 is subdivided into two equal halves with two different hopping
 parameters in such a way that the first half with $\kappa_1$ is in the
 symmetric phase, the second half with $\kappa_2$ in the Higgs phase.
 This means $\kappa_1 < \kappa_c < \kappa_2$.
 To be sure that this situation is stable we require that both halves
 stay in their phases at least for 20 autocorrelation times.
 Going to larger lattices $\kappa_{1,2}$ can be chosen closer to
 $\kappa_c$ at the order of inverse lattice volume.
 From the $2 \cdot 16^2 \cdot 128$ lattice at the {\em low} point and
 $2 \cdot 32^2 \cdot 256$ lattice at the {\em high} point the best
 estimates are, respectively,
\be \label{eq16}
\kappa_{c,L_t=2}^{low} = 0.12830(5) \ ,
\hspace{3em}
\kappa_{c,L_t=2}^{high} = 0.12887(1) \ .
\ee
 These are rather well reproduced, together with the volume dependence
 on smaller lattices, by the critical $\kappa$ obtained from the one
 loop gauge invariant effective potential of the length square
 of the scalar field $V_{eff}(\rho^2)$ \cite{LUSCHPOT,BUFOHE,FUTURE}.
 Therefore one can use $V_{eff}(\rho^2)$ for an estimate of
 $\kappa_{c,L_t=3,4,5,\cdots}$.
 By a quadratic interpolation of the $L_t=2,3,4$ critical points
 one obtains the following estimates for the derivatives of $\kappa$
 along the LCP's:
\be \label{eq17}
\frac{\partial\kappa^{low}}{\partial\tau} =-0.00064(6)           \ ,
\hspace{3em}
\frac{\partial\kappa^{high}}{\partial\tau} =-0.00111(13)         \ .
\ee

 The discontinuities in (\ref{eq14}) can be determined at the {\em low}
 point on large enough lattices directly at $\kappa=\kappa_c$, because
 the metastability is very strong and the configurations stay for a
 very long time in the phase they started from.
 At the {\em high} point a linear extrapolation can be used from
 the intervals $\kappa \leq \kappa_1$ and $\kappa \geq \kappa_2$, where
 no tunnelings to the other phase occur.
 On our $2 \cdot 16^2 \cdot 128$ and $2 \cdot 32^2 \cdot 256$ lattices
 at the {\em low} and {\em high} points, respectively, the finite volume
 corrections of order $(L_xL_yL_z)^{-1}$ are already smaller than
 the indicated statistical errors:
$$
\Delta L_\varphi^{low} =-21.84(4)                 \ , \hspace{2em}
\Delta Q_x^{low} =-643(2)                         \ , \hspace{2em}
\Delta P_U^{low} = 0.010557(15)                   \ ,
$$
\be \label{eq18}
\Delta L_\varphi^{high} =-1.026(12)                \ , \hspace{2em}
\Delta Q_x^{high} =-8.02(13)                       \ , \hspace{2em}
\Delta P_U^{high} = 0.000629(8)                    \ .
\ee
 Combining this with (\ref{eq14})-(\ref{eq15}) we arrive at
\be \label{eq19}
\left( \frac{\Delta\epsilon}{T_c^4} \right)^{low} = 1.68(17)   \ ,
\hspace{3em}
\left( \frac{\Delta\epsilon}{T_c^4} \right)^{high} = 0.125(19) \ .
\ee

 These results have been obtained on $L_t=2$ lattices alone.
 In the continuum limit $L_t \to \infty$, therefore later on one has to
 improve them by simulations on finer lattices with $L_t>2$
 \cite{FUTURE}.
 For a first qualitative comparison with two-loop resummed perturbation
 theory \cite{FODHEB} see fig. 1.


\section{Interface tension}
 Besides the latent heat another important physical quantity
 characterizing the strength of the phase transition is the
 interface tension $\sigma$ between the two phases, which can be
 determined on the lattice in several different ways.
 In the present letter we concentrate on the two-coupling
 method proposed by Potvin and Rebbi \cite{POTREB}, where an
 interface pair is enforced between the two halves of the periodic
 lattice.
 Since we have three bare parameters, in principle one can choose
 any of them (or some combination of them) to be different in the
 two halves.
 As in the latent heat the contribution of the $\varphi$-link
 $L_\varphi$ dominates, we decided to split the value of $\kappa$
 multiplying $L_\varphi$ in the action.
 On an $L_t \cdot L_x \cdot L_y \cdot L_z$ lattice with
 $L_z \gg L_{x,y,t}$, and in our case $L_t=2$, $L_x=L_y$, the
 two halves in the $z$-direction have hopping parameters $\kappa_{1,2}$
 which are slightly below and above the critical one:
 $\kappa_1 < \kappa_c < \kappa_2$.
 The interface tension between the states with $\kappa_1$ and
 $\kappa_2$ is given in lattice units by
\be \label{eq20}
a^3\sigma(\kappa_1,\kappa_2) = L_z \left\{
  \int_{\kappa_1}^{\kappa_2} d\kappa L_\varphi^{(1)}(\kappa,\kappa_2)
- \int_{\kappa_1}^{\kappa_2} d\kappa L_\varphi^{(2)}(\kappa_1,\kappa)
\right\} \ ,
\ee
 where $L_\varphi^{(1,2)}(\kappa,\kappa^\prime)$ denote the expectation
 values of the $\varphi$-link in the two halves if the hopping
 parameters are $\kappa$ and $\kappa^\prime$, respectively.

 In order to obtain the physically interesting interface tension one
 has, of course, to perform the limits $L_{x,y,z} \to \infty$ and
 $\Delta\kappa \equiv \kappa_2-\kappa_c = \kappa_c-\kappa_1 \to 0$.
 For a given lattice extension, however, $\Delta\kappa$ cannot be
 arbitrarily small, because if the difference in free energy density
 becomes small tunneling into the other phase can occur and the
 interfaces disappear.
 The presence of the interfaces can, however, be monitored to ensure
 the applicability of (\ref{eq20}).
 For small $\Delta\kappa$ the integrals in (\ref{eq20}) can be well
 approximated by the average of the values of the integrand at the two
 end points.
 Besides, for equal arguments we obviously have
 $L_\varphi^{(1)}(\kappa,\kappa) = L_\varphi^{(2)}(\kappa,\kappa)$.
 This gives
\be \label{eq21}
a^3\sigma \simeq L_z \Delta\kappa \left(
 L_\varphi^{(1)}(\kappa_1,\kappa_2) - L_\varphi^{(1)}(\kappa_1,\kappa_1)
+L_\varphi^{(2)}(\kappa_2,\kappa_2) - L_\varphi^{(2)}(\kappa_1,\kappa_2)
\right) \ .
\ee
 In the limit $\Delta\kappa \to 0$ the contributions with equal
 hopping parameters in the two halves do not contribute.
 The non-zero contributions can be determined by the expansion
$$
L_\varphi^{(1)}(\kappa_1,\kappa_2) = \frac{-c_1}{\kappa_1-\kappa_c}
+ b_1 + a_1(\kappa_1-\kappa_c) + O(\kappa_1-\kappa_c)^2 \ ,
$$
\be \label{eq22}
L_\varphi^{(2)}(\kappa_1,\kappa_2) = \frac{-c_2}{\kappa_2-\kappa_c}
+ b_2 + a_2(\kappa_2-\kappa_c) + O(\kappa_2-\kappa_c)^2 \ .
\ee
 This formula replaces (\ref{eq20}) in the region where
 $\sigma(\kappa_1,\kappa_2)$ depends linearly on $\kappa_{1,2}$.
 For $\Delta\kappa \to 0$ a finite volume estimator of the interface
 tension is
\be \label{eq23}
a^3\hat{\sigma} = L_z(c_1+c_2) \ .
\ee
 The advantage compared to (\ref{eq20}) is that no numerical evaluation
 of integrals is necessary.

 In our numerical simulations several lattice sizes up to
 $2 \cdot 16^2 \cdot 128$ at the {\em low} point and up to
 $2 \cdot 32^2 \cdot 256$ at the {\em high} point were exploited.
 In fact, the very long extensions in the $z$-direction have been chosen
 just for the purpose of also determining the interface tension.
 The measured $L_\varphi^{(1,2)}$ values were fitted in a carefully
 chosen $\kappa$-interval around the critical points in (\ref{eq16})
 by the three-parameter forms in (\ref{eq22}).
 At the {\em low} point we used the intervals
 $0.1279 \le \kappa_1 \le 0.1282$ and
 $0.1284 \le \kappa_2 \le 0.1287$ with seven points each.
 At the {\em high} point we used
 $0.12880 \le \kappa_1 \le 0.12886$ and
 $0.12888 \le \kappa_2 \le 0.12894$ with seven points each.
 The number of complete sweeps for measurements with an optimized ratio
 of heatbath to overrelaxation was between 20000 and 40000 per point.
 The obtained result from the fits in different subintervals is:
\be \label{eq24}
\left( \frac{\hat{\sigma}}{T_c^3} \right)^{low} = 0.84(16) \ ,
\hspace{3em}
\left( \frac{\hat{\sigma}}{T_c^3} \right)^{high} = 0.008(2) \ .
\ee
 Here $aT_c=\half$ has been used in order to convert from lattice to
 physical units.
 About half of the indicated errors come from statistics.
 The other half is a crude estimate of systematic errors, which
 is due to the choice of subintervals for $\kappa$.
 In general, choosing a subinterval closer to $\kappa_c$ gives smaller
 statistical errors and also somewhat smaller $c_{1,2}$.
 The quoted results are obtained from the more distant subintervals.

 This dependence on the chosen fit interval is a kind of rounding-off
 effect which is partly due to the interaction of the two interfaces.
 A similar dependence of $\sigma(\kappa_1,\kappa_2)$ on $\kappa_{1,2}$
 appears in (\ref{eq20}) if $\kappa_{1,2}$ are too close to $\kappa_c$.
 This $\kappa$-dependence is expected to be substantially reduced
 for $L_z \to \infty$ when the two interfaces do not interact.
 Of course, the transversal extensions $L_{x,y}$ also have to be taken
 to infinity.
 In our case comparisons to smaller lattices show in both points that
 within the given statistical errors $L_x=L_y$ are large enough.
 In principle, if $L_{x,y,z}$ are larger one can consider smaller
 $\Delta\kappa$ values, but this is beyond the scope of the present
 analysis.

 We also tried several other methods for the determination of $\sigma$.
 The results are in general qualitatively similar.
 A more detailed evaluation of the different methods will be
 included in a subsequent publication \cite{FUTURE}.
 A comparison of (\ref{eq24}) with two-loop resummed perturbation
 theory \cite{FODHEB} is shown in fig. 2.

\newpage

\section{Discussion}
 The outcome of our numerical studies on $L_t=2$ lattices is
 a qualitative agreement between two-loop resummed perturbation theory
 and four dimensional lattice simulations in the investigated range
 of Higgs boson masses, below 50 GeV.
 Note that at $M_H \simeq 18$ GeV the agreement between our data and
 perturbation theory becomes even better if, instead of $T_c$, $M_H$ is
 used to set the scale.
 Both latent heat and interface tension show that the first order
 symmetry restoring phase transition becomes substantially weaker
 for increasing Higgs boson mass (see eqs. (\ref{eq19}) and
 (\ref{eq24})).
 As it has been emphasized throughout this paper, although the $L_t=2$
 results may very well be qualitatively right, for a quantitative
 determination of the properties of the phase transition further
 numerical simulations on $L_t > 2$ lattices are necessary.


\vspace{1cm}
\large\bf Acknowledgements \normalsize\rm \newline
\vspace{3pt}

\noindent We thank our colleagues W. Buchm\"uller, A. Hebecker,
M. L\"uscher and R. Sommer for essential comments and proposals during
the course of this work.
F. Cs. and Z. F. were partially supported by Hung. Sci. Grant under
Contract No. OTKA-F1041/3-2190.


\vspace{2cm}
\begin{center}     \Large\bf Figure captions \normalsize\rm
\end{center}

\vspace{15pt}
\bf Fig.\,1.    \hspace{5pt} \rm
Comparison of the results of Monte Carlo simulations with two-loop
resummed perturbation theory \cite{FODHEB}.
The latent heat $\Delta\epsilon$ is shown as a function of the Higgs
boson mass.
The two curves correspond to the two slightly different gauge couplings
in (\ref{eq11}): the full line to the {\em low} value, the dashed line
to the {\em high} one.

\vspace{15pt}
\bf Fig.\,2.    \hspace{5pt} \rm
The same as fig. 1 for the interface tension $\sigma$.

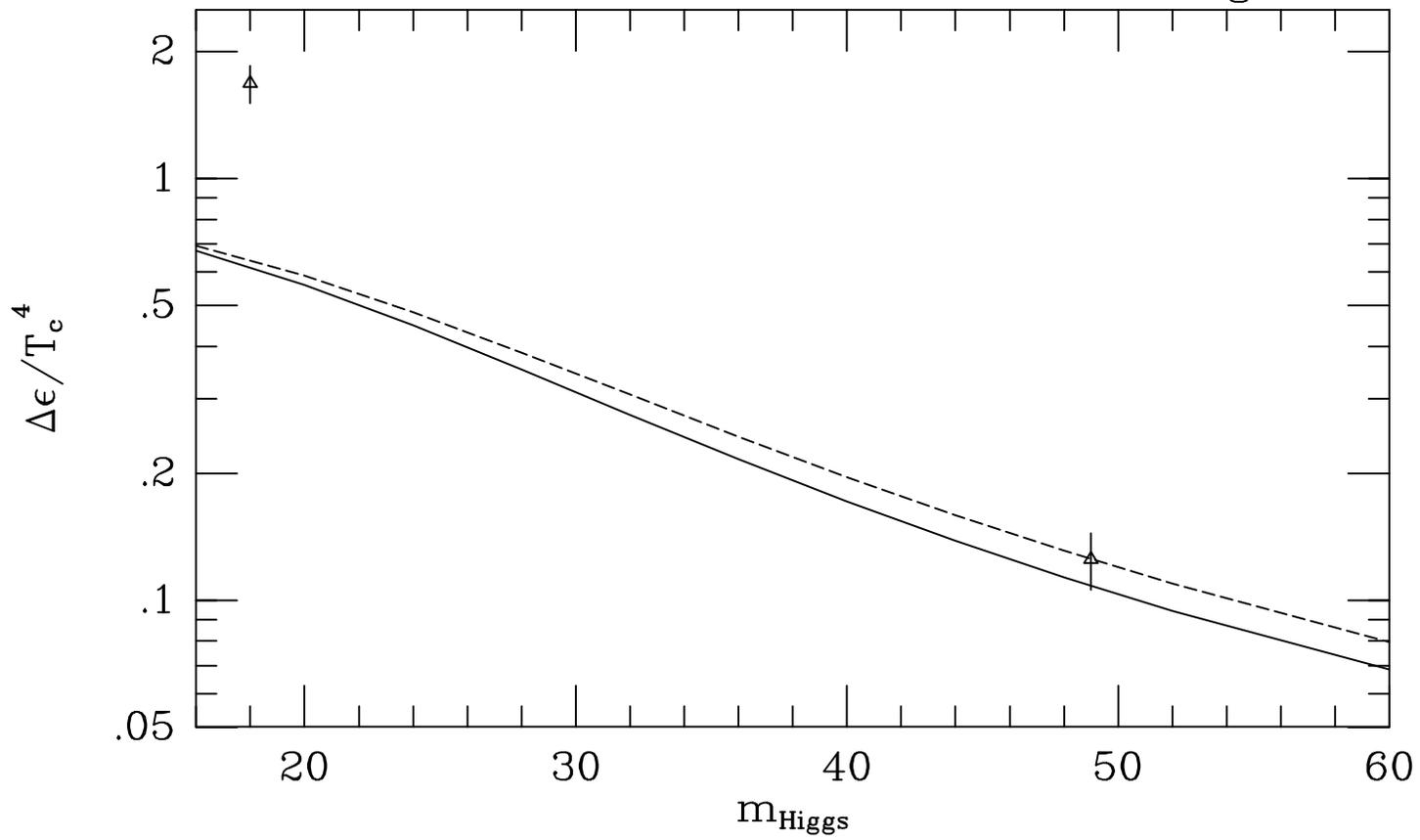
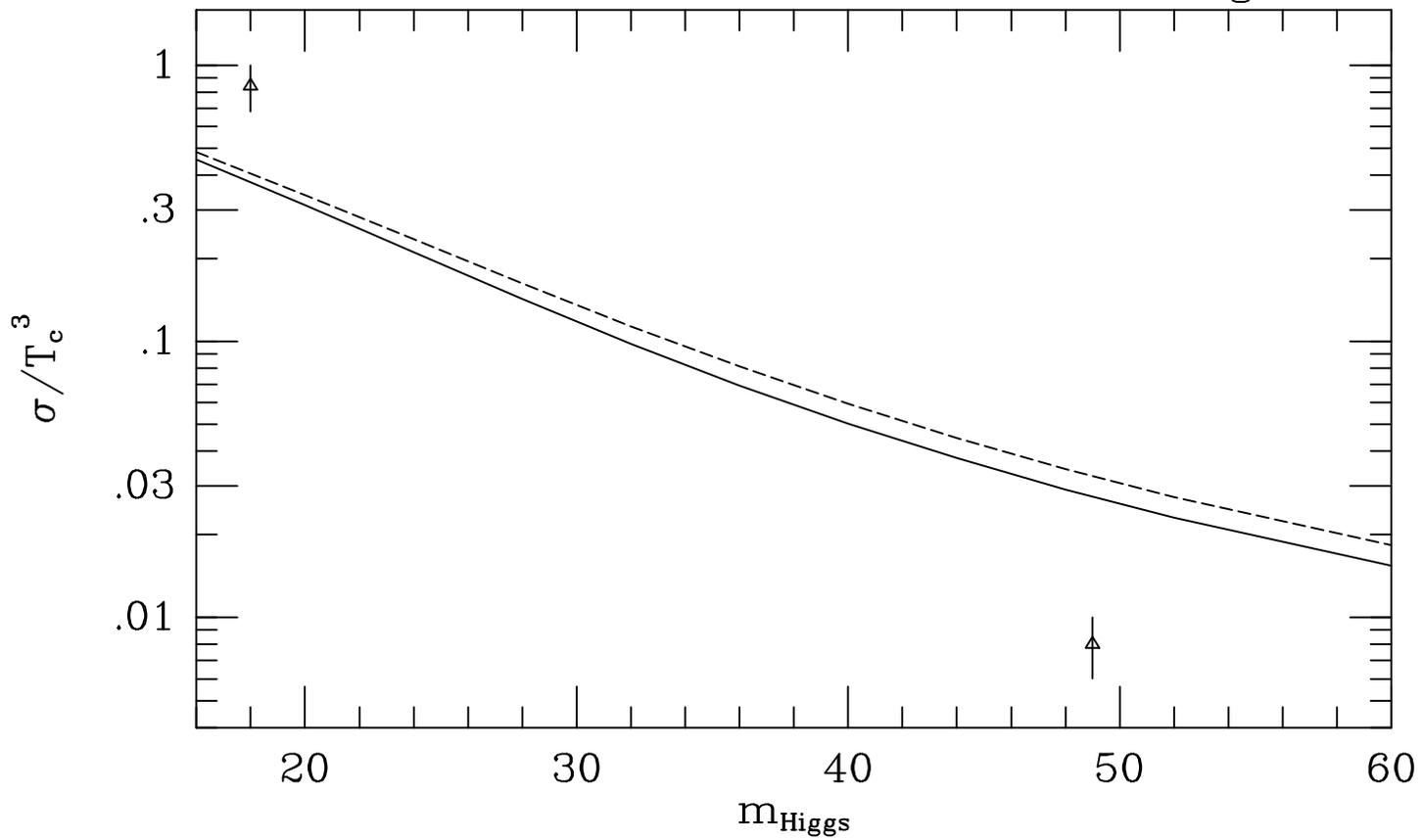
\end{document}